\documentclass[11pt,a4paper]{article}
\usepackage[T1]{fontenc}

\usepackage[a4paper,top=2cm,bottom=2cm,left=3cm,right=3cm,marginparwidth=1.75cm]{geometry}
 
\usepackage{amsmath}
\usepackage{graphicx}
\usepackage{hyperref}
\usepackage{algpseudocode}
\usepackage{algorithm}

\usepackage[affil-it]{authblk}

\makeatletter
\renewcommand{\section}{\@startsection {section}{1}{\z@}%
              {24pt}{12pt} {\large\scshape\bfseries}}

\renewcommand{\subsection}{\@startsection {subsection}{2}{\z@}%
             {12pt}{12pt}  {\itshape\bfseries}}

\usepackage{apacite}
\usepackage{natbib}

\title{\bfseries \normalsize Introducing a novel Location-Assignment Algorithm for Activity-Based Transport Models: CARLA}

\author[1]{Felix Petre*}
\author[1]{Lasse Bienzeisler}
\author[2]{Bernhard Friedrich}

\affil[1]{M.Sc., Institute of Transportation and Urban Engineering, Technische Universität Braunschweig, Germany}
\affil[1]{Univ.-Prof. Dr.-Ing., Institute of Transportation and Urban Engineering, Technische Universität Braunschweig, Germany}

\date{\vspace{-5ex}}

\begin{document}
\maketitle

\section*{Short summary}\small

This paper introduces CARLA (spacially Constrained Anchor-based Recursive Location Assignment), a recursive algorithm for assigning secondary (or any) activity locations in activity-based travel models. CARLA minimizes distance deviations while integrating location potentials, ensuring more realistic activity distributions. The algorithm decomposes trip chains into smaller subsegments, using geometric constraints and configurable heuristics to efficiently search the solution space. 

Compared to a state-of-the-art relaxation-discretization approach, CARLA achieves significantly lower mean deviations, even under limited runtimes. It is robust to real-world data inconsistencies, such as infeasible distances, and can flexibly adapt to various priorities, such as emphasizing location attractiveness or distance accuracy. 

CARLA’s versatility and efficiency make it a valuable tool for improving the spatial accuracy of activity-based travel models and agent-based transport simulations. 

Our implementation is available at \href{https://github.com/tnoud/carla}{https://github.com/tnoud/carla}.

\textbf{Keywords}: Activity-based model; agent-based; exhaustive search; location choice; relaxation– discretization algorithm; synthetic population

\section{Introduction}

Activity-based models (ABMs) have emerged as a more behaviorally realistic alternative to traditional demand modeling paradigms, such as the four-step model \citep{rasouliActivitybasedModelsTravel2014,bastariantoAgentbasedModelsUrban2023, rezvanyReviewActivitybasedDisaggregate2024}. 
Their disaggregate representation enables ABMs to better reflect individual decision-making processes and capture complex travel behavior patterns \citep{rasouliActivitybasedModelsTravel2014}. Thus, ABMs require modeling the attributes, intentions, and interactions of each individual traveler \citep{horlRelaxationDiscretizationAlgorithm2023}. 


A key challenge is the difficulty of robustly predicting activity location choices \citep{millerCurrentStateActivitybased2023}. Primary activities, such as commuting to work or attending educational institutions, can often be modeled with relative accuracy, supported by structured data sources like commuter matrices \citep{matetImprovingGenerationSynthetic2024}. Modeling the locations of secondary activities, including shopping, leisure, or dining, presents a significantly greater challenge \citep{matetImprovingGenerationSynthetic2024}. These decisions are influenced by several factors and exhibit variability, making them more complex to predict. Furthermore, the lack of standardized approaches and datasets exacerbates the difficulty of accurately representing these behaviors \citep{horlRelaxationDiscretizationAlgorithm2023}.

Common methods for the location choice of activities within activity plans use space-time prisms to define feasible activity locations, ensuring they are reachable within defined time or distant constraints \citep{salvadeRepresentingLocationChoice2022}. Another approach involves the use of mobile phone data, which can be utilized to generate trip schedules for activity-based traffic simulations, enabling the estimation of global spatial mobility behavior of a population \citep{cuiGeneratingSyntheticProbabilistic2021, matetImprovingGenerationSynthetic2024}.

\cite{horlRelaxationDiscretizationAlgorithm2023} introduced a data-driven algorithm for assigning secondary activity locations, ensuring consistency with fixed points in daily activity plans while maintaining realistic distance distributions. This method eliminates the need for complex choice models and supplementary origin-destination data, providing a practical and user-friendly solution for secondary location assignment.
The algorithm has proven effective in various ABMs \citep{horlSyntheticPopulationTravel2021, horlSimulationPriceCustomer2021,sallardOpenDatadrivenApproach2021,pereiraAdvancedTravelDemand2022,manoutImplicationsPricingFleet2024} and represents the current state-of-the-art. 

Although computationally efficient, the approach has potential for further improvement. The algorithm iteratively selects a feasible option from the solution space to find a valid result, but it does not guarantee the optimal solution. A more systematic optimization algorithm could reliably achieve the best possible outcome. Additionally, the algorithm does not account for \textit{location potential}, i.e. the prominence or attraction of a certain location. As a result, larger destinations, such as shopping malls, are not selected more frequently than smaller shops, limiting the representation of realistic activity patterns.



In this work, we present a new algorithm for assigning secondary activity locations, or any activity chain, using Euclidean distances from a household travel survey (HTS). The algorithm utilizes a recursive strategy to decompose activity chains into manageable segments and systematically narrows the solution space using tailored heuristics and geometric constraints.
It incorporates location potentials, ensuring realistic assignments. 
Its flexibility allows it to adapt to various scenarios and priorities; for example, when location potentials are more important, the algorithm can prioritize them while using distances solely to define the search space. Furthermore, the algorithm is robust and capable of handling implausible distance inputs without significant degradation in performance.



\section{Methodology}

\begin{figure}
    \centering
    \includegraphics[width=0.75\linewidth]{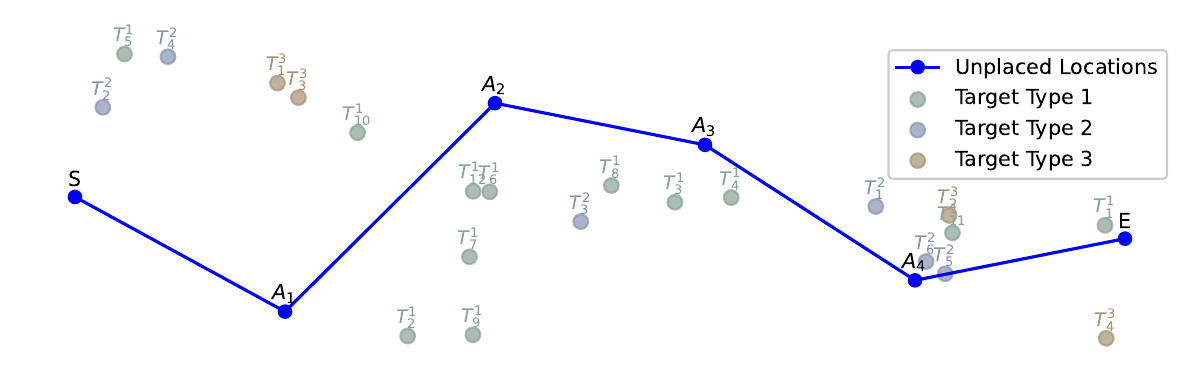}
    \caption{A five-trip activity chain with some illustrated target locations.}
    \label{fig:chain_targets}
\end{figure}

Building on the conceptual framework established by \cite{horlRelaxationDiscretizationAlgorithm2023}, we adopted a similar terminology to describe the activity assignment problem. An activity chain represents a sequence of activities performed by an individual, connected by trips, which denote the travel between consecutive activities. Within an activity chain, illustrated in \autoref{fig:chain_targets}, a main activity (usually \emph{work} or \emph{education}) is placed using a different algorithm, leaving secondary activities to be located.
Activity chains are divided into segments: a segment is a sequence of trips with known locations at its start and end and any number of activities with unknown locations in between. 


For each segment bounded by two fixed locations \( S \) (start) and \( E \) (end), the goal was to assign \( n \) variable activities \( A_1, A_2, \dots, A_n \) (e.g. sport, leisure...) to discrete target locations \( T_1, T_2, \dots, T_k \) from a predefined set of locations in the study area (e.g. parks, businesses...), where the desired activity type must fit the target location type.
The problem can be divided into two cases:

\begin{enumerate}
    \item \textbf{Two-trip Case}: This involves a single variable activity \( A \) located between the two fixed points \( S \) and \( E \). It is relatively straightforward to solve.

    \item \textbf{Complex case}: This involves \( n \geq 2 \) variable activities \( A_1, A_2, \dots, A_n \) distributed between \( S \) and \( E \). The solution space becomes extremely large and complex as \( n \) increases, creating a highly interdependent optimization problem. Each activity placement \( T_j \) affects the subsequent trips.
\end{enumerate}

The objective was to minimize the total discrepancy between the distances \( d_{ij}^{\text{HTS}} \) reported in the HTS and the distances \( d_{ij}^{\text{model}} \) derived from the assigned locations. In order to incorporate location potentials \( P(T_j) \) for the target candidate locations \( T_j \) into the optimization objective, a weighted potential term was added:

\begin{equation}
\text{Maximize} \quad \alpha \sum_{j=1}^k P(T_j) - \beta \sum_{i=1}^{n+1} \left| d_{i-1,i}^{\text{HTS}} - d_{i-1,i}^{\text{model}} \right|
\end{equation}

where \(d_{i-1,i}^{\text{HTS}}\) represents the distance between consecutive activities as observed in the HTS, \(d_{i-1,i}^{\text{model}}\) represents the Euclidean distance between their assigned locations in the models, and \( \alpha \) and \( \beta \) are tunable parameters controlling the influence of potentials and deviations.


\subsection{Core Idea}
To address the optimization problem, we developed a recursive algorithm that simplifies the trip chain step-by-step until a solution is found. The key idea is to divide a complex segment into two smaller subsegments at a middle activity, the \emph{anchor}, which is placed at a discrete, suitable location. This splits the segment \((S \rightarrow E)\) into \((S \rightarrow A_m)\) and \((A_m \rightarrow E)\). The process is repeated for all subsegments until all activities are assigned. A tree search evaluates multiple anchor placements and subsegments simultaneously. Each \textit{branch} is scored based on distance deviations and location potentials, selecting the best solution. Geometric constraints narrow the search space, and selection heuristics improve efficiency.

\subsection*{Key Components}

\textbf{1. Main Algorithm Loop }

For each person, the algorithm processes each segment sequentially in the main loop. The results for each segment are combined to form a complete, placed trip chain.


\vspace{1em}
\textbf{2. SolveSegment Function}

The \texttt{SolveSegment} function (see Algorithm \ref{alg:solve_segment}) is the core of the algorithm. It operates on the given segment by selecting a number of possible anchor locations (\emph{candidates}).
Each chosen anchor spawns two new subsegments, unique to this anchor location, which are solved recursively.
The algorithm proceeds until it reaches one of two base cases: 
\textit{(i)} a trivial single-trip subsegment (where no further activity placement is needed), or 
\textit{(ii)} the \textit{Two-trip Case} (where the best activity location can be found directly).
It returns the best identified segment placement.




\begin{algorithm}
\caption{SolveSegment\label{alg:solve_segment}}
\begin{algorithmic}[1]
\Require Segment $\text{seg}$, target locations $T$, configuration $C$
\Ensure Optimally placed segment $\hat{\text{seg}}$, score
\If{$|\text{seg}| = 1$} \Comment{Base case: single trip}
    \State \Return $\text{seg}, 0$ 
\ElsIf{$|\text{seg}| = 2$} \Comment{Base case: two trips}
    \State $\text{candidates} \gets \text{CircleIntersections}(\text{seg}[0].\text{from}, \text{seg}[1].\text{to}, T, C)$
    \State $\text{scores} \gets \text{EvaluationFunction}(\text{candidates})$
    \State $\text{selected\_candidate, selected\_score} \gets \text{SelectionFunction}(\text{candidates, scores, 1})$
    \State Update $\text{seg}$ with $\text{selected\_candidate}$
    \State \Return $\text{seg}, \text{selected\_score}$
\Else \Comment{Recursive case: multiple trips}
    \State $\text{anchor} \gets \text{ChooseAnchor}(\text{seg}, C)$
    \State $D_1min, D_1max, D_2min, D_2max \gets \text{GetFeasibleDistances}(\text{seg}, \text{anchor})$ 
    \State $\text{candidates} \gets \text{OverlappingRings}(D_1min, D_1max, D_2min, D_2max, T, C)$
    \State $\text{scores} \gets \text{EvaluationFunction}(\text{candidates})$
    \State $\text{selected\_candidates}, \text{selected\_scores} \gets$ \\
    \hspace{2em}$\text{SelectionFunction}(\text{candidates}, \text{scores}, C.\text{number\_of\_branches})$
    \State $\text{branch\_segments}, \text{branch\_scores} \gets []$
    \ForAll{$c \in \text{selected\_candidates}$}
        \State Update $\text{seg}$ with candidate $c$ at $\text{anchor}$
        \State $\hat{\text{seg}}_1, \text{score}_1 \gets \text{SolveSegment}(\text{seg}[0:\text{anchor}], T, C)$
        \State $\hat{\text{seg}}_2, \text{score}_2 \gets \text{SolveSegment}(\text{seg}[\text{anchor}+1:], T, C)$
        \State Combine $\hat{\text{seg}}_1$ and $\hat{\text{seg}}_2$ into $\hat{\text{seg}}$
        \State $\hat{\text{seg}}\text{\_score} \gets \text{score}_1 + \text{score}_2 + \text{score of candidate $c$}$
        \State $\text{branch\_segments.append}(\hat{\text{seg}})$
        \State $\text{branch\_scores.append}(\hat{\text{seg}}\text{\_score})$
    \EndFor
    \State $\text{best\_index} \gets \arg\max \text{branch\_scores}$
    \State \Return $\text{branch\_segments[best\_index]}, \text{branch\_scores[best\_index]}$
\EndIf
\end{algorithmic}
\end{algorithm}

In the \textit{Two-trip Case}, ideal possible locations for \( A \) are at the intersections of the two circles:
\begin{itemize}
    \item Centered at \( S \) with radius \( d_{S,A} \),
    \item Centered at \( E \) with radius \( d_{A,E} \).
\end{itemize}
If the input distances are infeasible (i.e., the circles do not intersect because they are too far apart or one being fully enclosed within the other), the algorithm selects a single point that minimizes total deviation.

\begin{figure}
    \centering
    \includegraphics[width=1\linewidth]{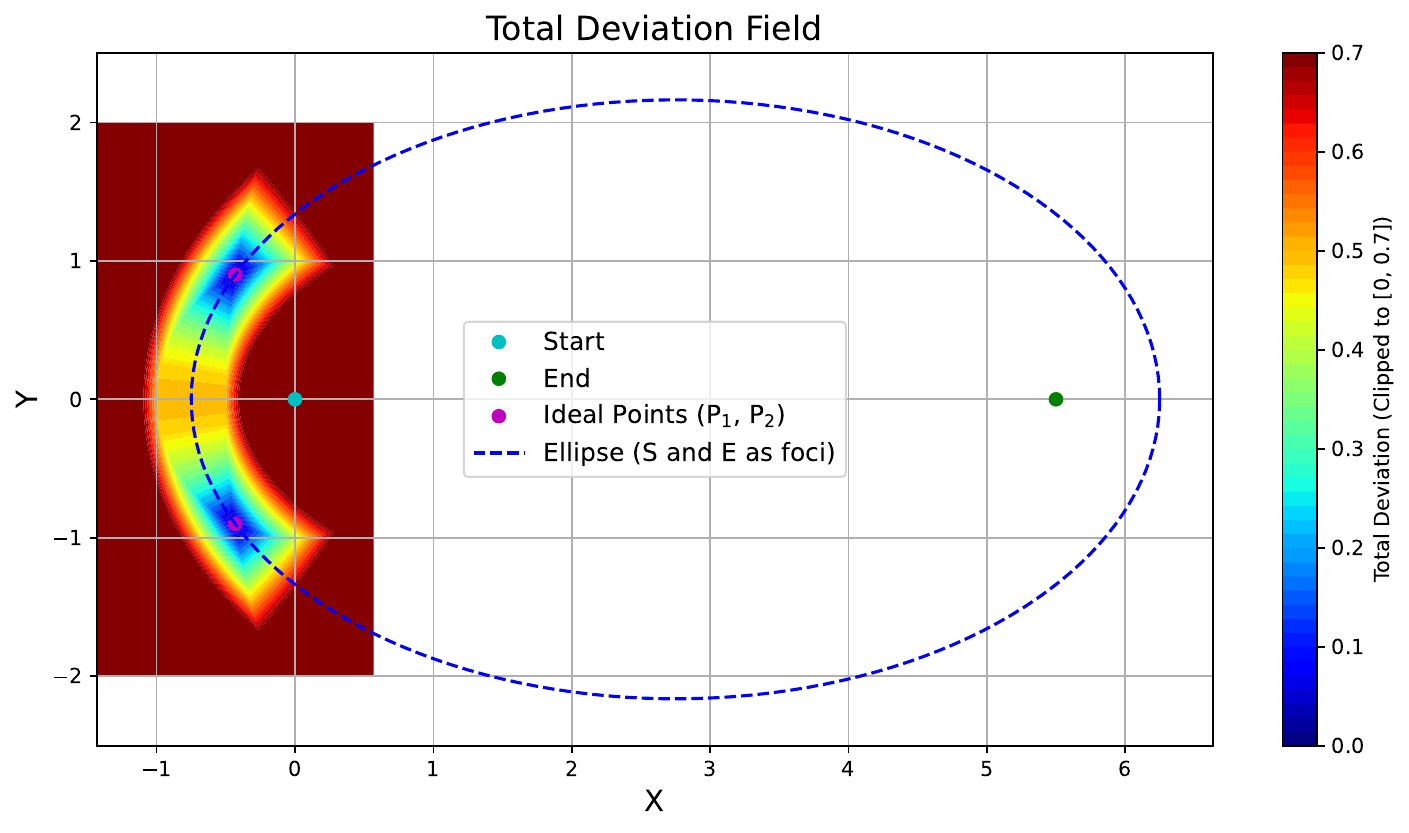}
    \caption{Total distance deviation of locations around the ideal points}
    \label{fig:total_deviation_field}
\end{figure}

After determining the ideal point(s), the best discrete location is chosen from the target locations. This, however, is not necessarily the location closest to an ideal point. First, when location potentials are taken into account, a location further from an ideal point but with a much larger potential may be superior. Additionally, the Euclidean proximity to an ideal point is not directly representative of the total distance deviation, shown in this example: Given an expected distance of 1 unit from \( S \) and 6 units from \( E \), the ideal points are found as illustrated in \autoref{fig:total_deviation_field}. 
Moving orthogonally to the illustrated ellipse from either ideal point has a stronger influence on the distances from \( S \) and \( E \) than moving tangentially to the ellipse by the same distance.
The deviation gradients around the ideal points do not form circular funnels, but a more complex shape.

Thus, instead of choosing the closest location to an ideal point, we select a number of candidates around each ideal point, which are then evaluated (by the \texttt{Evaluation Function}) and the single best candidate selected (by the  \texttt{Selection Function}). The difference between choosing the closest location and the "best" among several close locations is significant when optimizing for minimal deviation. 


\begin{figure}
    \centering
    \includegraphics[width=0.6\linewidth]
    {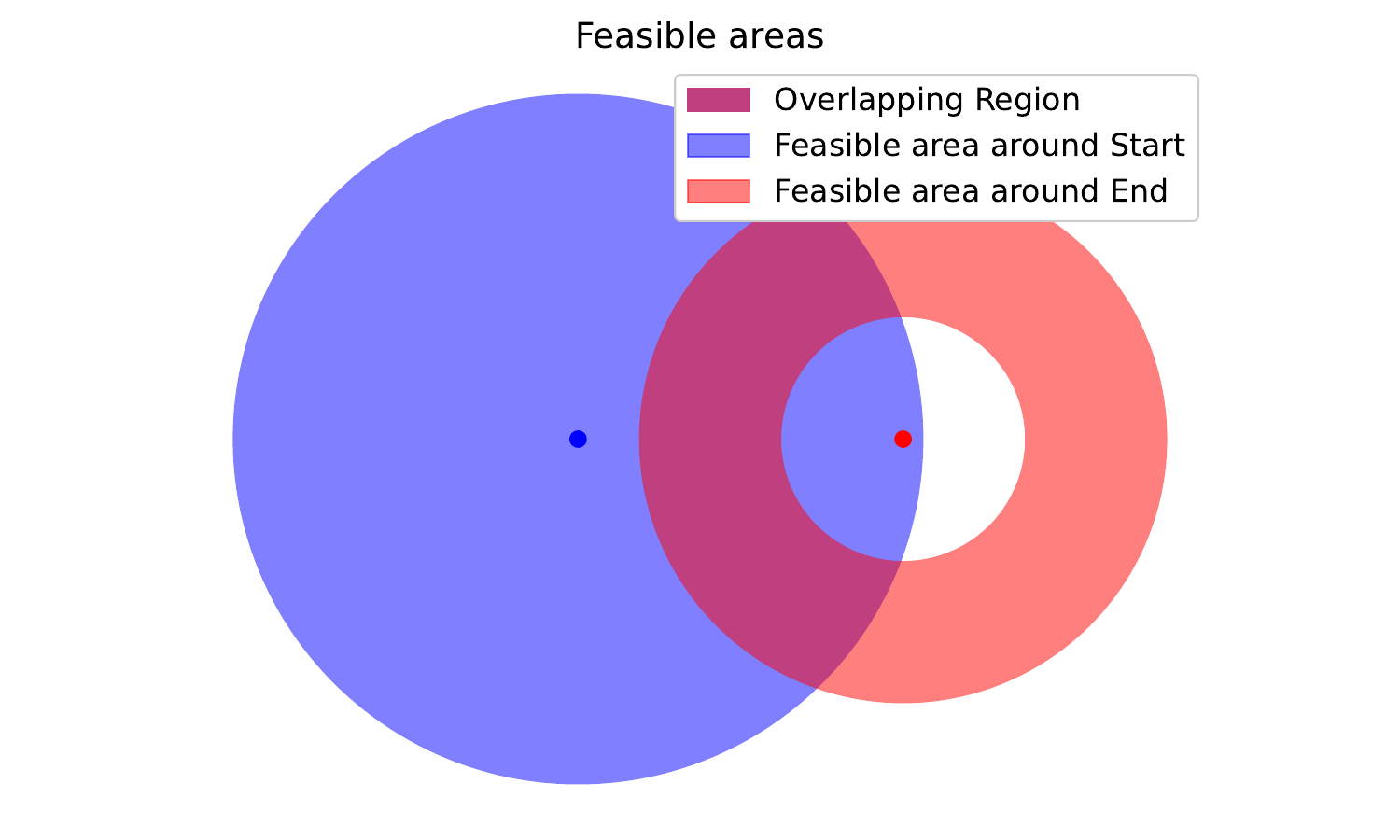}
    \caption{Feasible areas for the anchor location around Start and End}
    \label{fig:feasible_areas}
\end{figure}

In the \textit{Complex, recursive, Case}, the ideal anchor location is highly dependent on both previous and following trips, whose locations are yet undetermined. Using geometric constraints, the algorithm restricts the search area to overlapping feasible regions (rings, \autoref{fig:feasible_areas}) around the start (\( S \)) and end (\( E \)) points, defined by:
\begin{itemize}
    \item Inner radius (\(r_{\text{min}}\)): the minimum allowable distance.
    \item Outer radius (\(r_{\text{max}}\)): the maximum allowable distance.
\end{itemize}

The maximum radii are the sum of all trip distances between \( S \) and the anchor (\( A_m \)), and \( A_m \) and \( E \), respectively. Minimum radii are calculated combinatorially as the shortest distance reachable from \( S \) or \( E \), considering trip lengths. For example, a very long trip followed by two short trips may prevent returning to the reference point, making \( r_{\text{min}} \) the difference between the long trip and the combined shorter trips.
By constraining the search space to the overlap of these rings, the algorithm ensures that the placement respects travel distance constraints while significantly reducing computational effort. 

Once the feasible region is determined, the function queries the spatial index for all locations within it which fit the given activity type (\textit{candidates}). If there are no candidate locations found, this either implies that the rings do not overlap indicating an impossible trip chain (and may be expected given the nature of HTS), or simply indicates that there are no acceptable target locations in the overlap. In either case, the rings are expanded by shrinking the minimum and increasing the maximum radii until a specified minimum number of valid location is found. 
The identified candidates are then evaluated using the \texttt{Evaluation Function} and a subset of configurable size is selected using the \texttt{Selection Function}, ensuring a balance between location potential and distance deviations. The selected candidates then serve as inputs for the subsequent recursive step.

\vspace{1em}
\textbf{3. Scoring and Selection}



The \texttt{Evaluation Function} assigns a score to each candidate:

\begin{equation}
\text{Score}(c) = \alpha \cdot P(c) - \beta \cdot D(c)
\end{equation}

Here, \( \alpha \) and \( \beta \) are weights that control the trade-off between location potential \( P(c) \) and the distance deviation \( D(c) \). This formulation can be extended to include nonlinear terms if needed.

For the Two-trip Case, \( D(c) \) is defined as:

\begin{equation}
D(c) \;=\; \bigl|\,d_{1} \;-\; \|S - c\|_2 \bigr|
\;+\;
\bigl|\,d_{2} \;-\; \|c - E\|_2 \bigr|.
\end{equation}

Here, \( d_1 \) and \( d_2 \) are the expected distances, while \( \|S - c\|_2 \) and \( \|c - E\|_2 \) are the Euclidean distances between the points. In the one-trip case, this simplifies to a single term.

The final score of a whole chain is computed as the sum of scores for all selected locations in the chain. By summing up the individual contributions of spatial potential and distance accuracy, this resolves directly into the optimization function introduced earlier.




The \texttt{Selection Function} is responsible for selecting the best candidate locations based on their scores as computed by the Evaluation Function. While selecting more candidates (and thus creating more branches) improves the quality of the results, it also significantly increases processing times, particularly for long activity chains. To address this trade-off, the function supports several configurable heuristic strategies, allowing the algorithm to balance exploration, diversity, and computational efficiency. They are listed in \autoref{tab:selection_strategies_summary}.

\begin{table}[h!]
\centering
\caption{Selection Strategies Overview}
\label{tab:selection_strategies_summary}
\begin{tabular}{p{0.30\linewidth} p{0.65\linewidth}}
\hline
\textbf{Strategy} & \textbf{Description} \\
\hline
\textbf{Keep All} & Selects all candidates, ensuring the optimal solution but with high computational cost. \\
\textbf{Top-\( k \)} & Retains the \( k \) highest-scoring candidates, effective but may overlook interdependent solutions. \\
\textbf{Monte Carlo} & Selects candidates probabilistically based on normalized scores, introducing exploration and stochasticity. \\
\textbf{Top-\( k \) Monte Carlo} & Combines deterministic (\( k \) top candidates) and probabilistic sampling, balancing quality and diversity. \\
\textbf{Spatial Downsampling} & Selects candidates evenly from a configurable grid of cells to ensure distribution across the feasible region. Much faster than the alternative of using k-means clustering.\\
\textbf{Top-\( k \) Spatial Downsampling} & A hybrid approach: selects \( k \) top candidates and applies spatial downsampling for diversity. \\
\hline
\end{tabular}
\end{table}

\section{Results and discussion}

\subsection{Comparing Hörl and CARLA: Trade-offs Between Performance and Accuracy}

We evaluated the trade-off between runtime and accuracy for both Hörl and CARLA algorithms. For Hörl, we varied the maximum number of iterations, while for CARLA, we varied the number of branches. Since Hörl minimizes distance deviations only, we simplified CARLA's Evaluation Function to:
\begin{equation}
\text{Score}(c)
=
-\,\text{D}(c)
\end{equation}

We profiled Hörl's algorithm and found no evident bottlenecks in its implementation. 
The evaluation used a sample of 1000 individuals from a German HTS, prefiltering persons with trips over 30 km as they exceeded the study area's bounds. Target locations were real-world data from Hanover, Germany, with activity types derived from OpenStreetMaps.

Hörl’s implementation uses a relaxation solver and an assignment solver, with the maximum number of iterations set to 1000 and 20, respectively. 
Deviation thresholds determine when the algorithm returns. While they improve runtime, they limit the best achievable result. To evaluate this, we tested two configurations: the standard setup and a testing setup with increased limits and reduced thresholds, as shown in \autoref{tab:hoerl_comparison}.
\begin{table}[h!]
  \begin{center}
    \caption{Comparison of Hörl Standard and Hörl 1m1000 Configurations.}
    \vspace{2ex}
    \label{tab:hoerl_comparison}
    \begin{tabular}{l|l|l} 
      \textbf{Parameter} & \textbf{Hörl Standard} & \textbf{Hörl 1m1000} \\
      \hline
      Thresholds (Car driver/passenger, Public transport) & 200 m & 1 m \\
      Thresholds (Walk, Bike) & 100 m & 1 m \\
      AssignmentSolver Max Iterations & Limited to 20 & Limited to 1000 \\
    \end{tabular}
  \end{center}
\end{table}

Evaluating CARLA's perfomance, we applied a consistent base configuration (see Table~\ref{tab:advanced_petre_config}). 

\begin{table}[b!]
  \begin{center}
    \caption{Base CARLA Configuration.}
    \vspace{2ex}
    \label{tab:advanced_petre_config}
    \begin{tabular}{l|l} 
      \textbf{Parameter} & \textbf{Value} \\
      \hline
      number\_of\_branches & 50 \\
      min\_candidates\_complex\_case & 10 \\
      candidates\_two\_trip\_case & 20 \\
      anchor\_strategy & lower\_middle \\
      selection\_strategy\_complex\_case & mixed (Top-\( k \) Monte Carlo Sampling) \\
      selection\_strategy\_two\_trip\_case & top\_k \\
    \end{tabular}
  \end{center}
\end{table}

Before using the relaxation solver, Hörl's algorithm polls distances from a distribution, as its input data lack direct distance values. In contrast, we directly input Euclidean distances, bypassing this step. However, Hörl's relaxation solver relies on feasible distances and performs poorly when handling infeasible ones. To address this, we tested two scenarios: one using the full HTS sample and another excluding persons with infeasible data, leaving 693 individuals for analysis. 

Both algorithms can either be applied to locate all activities of a activity chain or used only between already assigned fixed main activities. Accordingly, we evaluated two additional scenarios: one where the main activity is pre-placed at a fixed location, requiring only the segments before and after the main activity to be placed, and another where the main activity is not pre-placed, requiring the entire activity chain to be assigned. In sum, the four scenarios emerged:

\begin{enumerate}
    \item Main activity pre-placed, considering only feasible trip chains.
    \item Main activity pre-placed, considering all trip chains.
    \item Main activity not placed, considering only feasible trip chains.
    \item Main activity not placed, considering all trip chains.
\end{enumerate}

\begin{figure}
    \centering
    \includegraphics[width=1\linewidth]{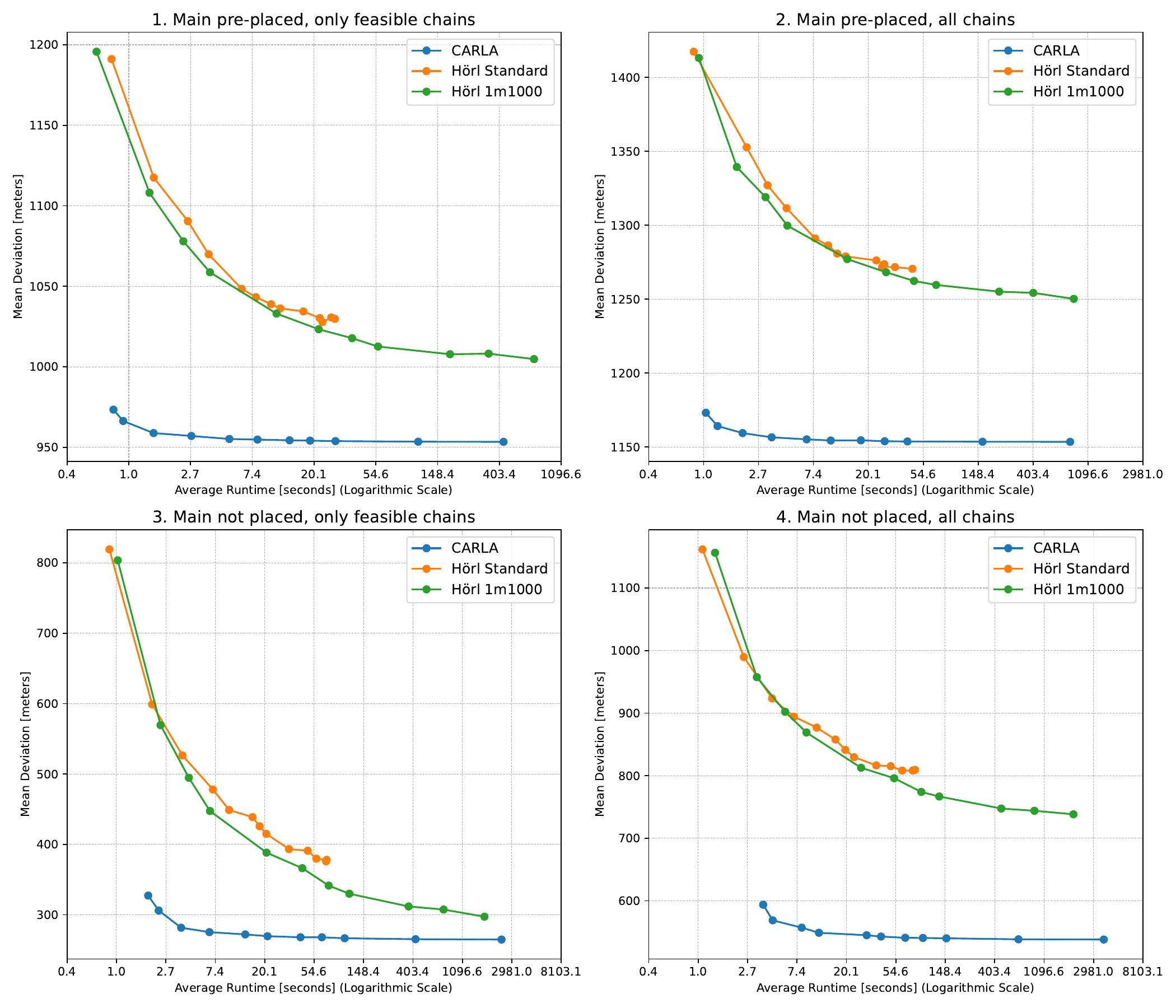}
    \caption{Mean deviation vs. runtime across scenarios, varying iterations for Hörl and branches for CARLA. Each point in the plots represents a result of a distinct parameter configuration of the respective algorithm.}
    \label{fig:results}
\end{figure}


The results in \autoref{fig:results} showcase the performance of both algorithms. CARLA consistently outperforms Hörl in all scenarios, achieving solutions with a mean deviation close to the likely optimum. This is even the case at much shorter runtimes, with improvements observed as the number of branches increased. With a low number of branches (20), CARLA's results are not only better but also significantly faster.

\subsection{Detailed comparison}

A single run from scenario three, which excludes infeasible chains to isolate the algorithms' performance, highlights the differences. Using its base settings (\autoref{tab:advanced_petre_config}), CARLA completed processing in 45 seconds, while Hörl's standard configuration, with a maximum of 1000 iterations as per the published standard, required 89 seconds.

\begin{figure}
    \centering
    \includegraphics[width=1\linewidth]{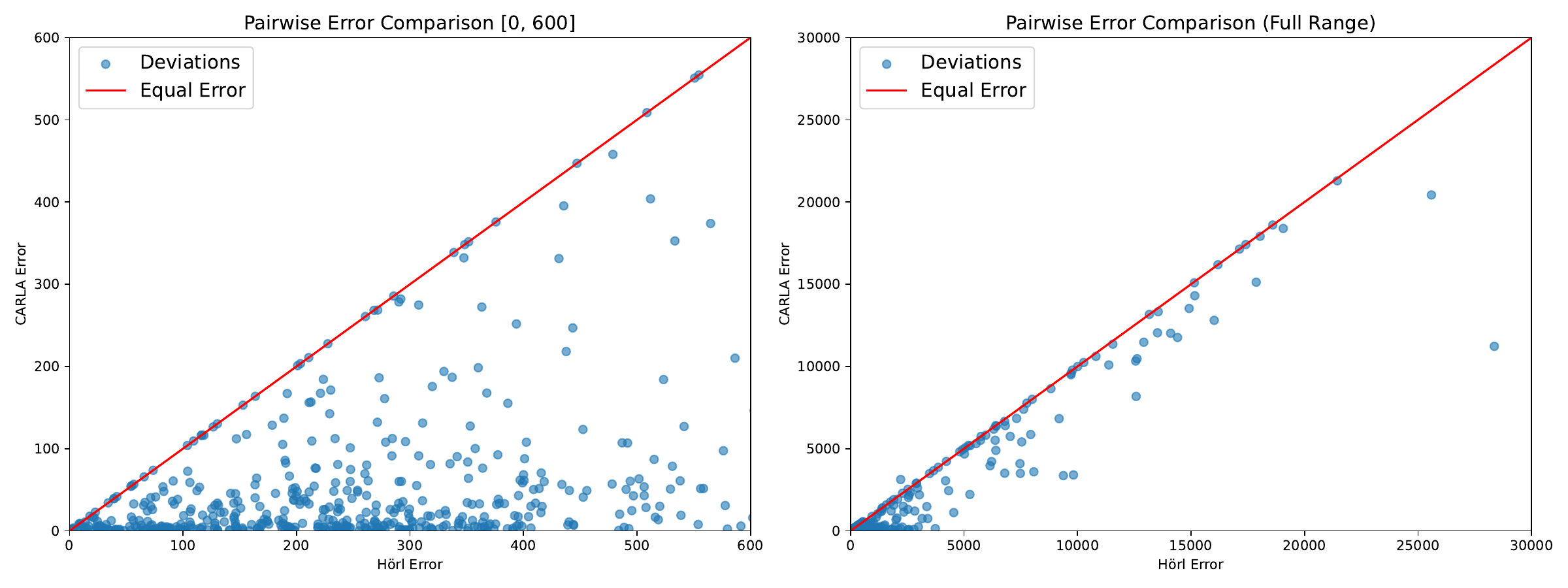}
    \caption{Pairwise comparison of total distance deviation per person.}
    \label{fig:detailed_comparison}
\end{figure}

CARLA produced better results in 630 cases (90.9\%). In 62 cases (8.9\%), results were equal, indicating that both algorithms chose the same discrete locations. In a single case, Hörl found a location chain with a smaller total deviation, reflecting the heuristic nature of CARLA's settings in this test. 

The results are visualized in \autoref{fig:detailed_comparison}. As 92.5\% of total deviations produced by the Hörl algorithm are below 600m, we focus on this area first. CARLA consistently manages to find total deviations well below Hörl's deviations, very often close to zero.
The full range indicates that even for chains where a placement close to the given distances is not possible (e.g. due to sparse target locations), CARLA finds better results.

\section{Conclusions}
This paper presented CARLA, a novel recursive algorithm for assigning (secondary, or any) activity locations in activity-based travel models. CARLA optimizes for both location potential and adherence to observed distance distributions while handling real-world complexities such as infeasible distances or sparse target locations.
Compared to Hörl’s relaxation-discretization algorithm, CARLA consistently achieved lower mean deviations, even under constrained runtimes. Its recursive structure efficiently decomposes trip chains into manageable segments, leveraging geometric constraints to reduce the solution space.
Configurable evaluation and selection functions allow CARLA to balance computational efficiency with solution quality and enable the integration of location potentials for improved realism.
While CARLA demonstrates robustness and flexibility, its performance ultimately depends on accurate HTS inputs. Future work will focus on integrating it into a full framework, where it may replace several existing algorithms.

\section*{Acknowledgements}
\noindent\parbox{\textwidth}{%
The results were created as part of the Urban Climate Future Lab UCFL, funded by zukunft.niedersachsen, a funding program of the Lower Saxony Ministry of Science and Culture and the Volkswagen Foundation. Responsibility for the content of this publication lies with the authors.
}

\bibliography{references}

\end{document}